\documentclass[12pt]{article}

\usepackage[totalheight = 23cm, totalwidth = 17cm]{geometry}
\usepackage{amssymb,amsmath,amsfonts,amsbsy,graphicx,bm}

\usepackage{color,cancel,ulem, hyperref}

\newcommand{\dis}[1]{\begin{equation}\begin{split}#1\end{split}\end{equation}}

\begin{document}

\begin{titlepage}

\begin{center}

{\setlength{\baselineskip}%
{1.5\baselineskip}
{\LARGE \bf 
 Species scale associated with Weinberg operator and bound on Majorana neutrino mass
}
\par}

\vskip 1.0cm

{\large
Min-Seok Seo$^{a}$ 
\footnote[0]{e-mail : minseokseo57@gmail.com}
}

\vskip 0.5cm

{\it
$^{a}$Department of Physics Education, Korea National University of Education,
\\ 
Cheongju 28173, Republic of Korea
}

\vskip 1.2cm

\end{center}

\begin{abstract}

 When states in a tower like the Kaluza-Klein or the string tower couple to another state  through the irrelevant operators of the same type, their contributions to the loop corrections of the relevant or the marginal operators are not negligible,   threatening the perturbativity.
 This can be avoided provided the cutoff scale is lower than the species scale associated with the irrelevant operator.
 We apply this to towers of states associated with the neutrino which couple to the Higgs through the Weinberg operator, the dimension-5 irrelevant operator generating the Majorana neutrino mass.
 Requiring the `Majorana species scale', the species scale associated with the Weinberg operator, to be below the gravitational species scale, one finds the lower bound on the Majorana neutrino mass determined by the species number.
 The Festina-Lente bound also gives the lower bound on the Majorana neutrino mass, but it is not so stringent.
 Meanwhile, even if the neutrino mass is of the Dirac type at the renormalizable level, the Majorana mass term still can be written in the effective field theory action so far as the Weinberg operator is not forbidden.
 Even if the Majorana neutrino mass is larger than the Dirac one, so far as there are sufficient degrees of freedom with mass smaller than the scale of the cosmological constant,  the observation of the Majorana nature of the neutrino may not contradict to quantum gravity constraints which rules out  the neutrino mass purely given by the Majorana type.

\end{abstract}

\end{titlepage}

\newpage

\section{Introduction}

Explaining nonzero but extremely tiny neutrino mass is one of long-standing puzzles in particle physics (for reviews, see, e.g., \cite{ParticleDataGroup:2024cfk} and also \cite{Fukugita:2003en, Giunti:2007ry}).
As a guiding principle to address this problem, the technical naturalness has been considered, according to which the small parameter in the low energy effective field theory (EFT) is natural if an enhanced symmetry develops in the vanishing limit of the parameter \cite{tHooft:1979rat}.
Regarding the nature of the neutrino mass, the technical naturalness criterion seems to prefer the Majorana type to  the Dirac one,  since the smallness of the former is protected by the lepton number symmetry.
Indeed, in the standard model (SM) of particle physics, the   leading interaction that violates the lepton number symmetry comes from a dimension-5 irrelevant operator $(l\cdot h)^2/M$ called the Weinberg operator, where $l$ and $h$ indicate the lepton and the Higgs doublets, respectively \cite{Weinberg:1979sa}.
When the Higgs acquires the vacuum expectation value (VEV) $v$, it induces the Majorana neutrino mass $v^2/M$, which is suppressed by $v/M$ compared to the electroweak scale $v$.
The well known UV completion of the Weinberg operator might be the type I see-saw mechanism \cite{Minkowski:1977sc, Yanagida:1979as, Gell-Mann:1979vob, Glashow:1979nm, Mohapatra:1979ia, Schechter:1980gr}.
 In this scenario, the Majorana fermion $\nu_R$ called the right-handed neutrino is introduced in addition to the SM particles, which couples to $l$ and $h$ through the Yukawa interaction $(l\cdot h)\nu_R$.
 When $\nu_R$ carries the lepton number, the lepton number symmetry can be imposed on  the Yukawa interaction, but it is broken by the Majorana mass term of $\nu_R$.
 The  Weinberg operator is obtained by integrating out $\nu_R$ then $1/M$ is given by (Yukawa coupling)$^2$/(Majorana mass of $\nu_R$).
For  ${\cal O}(1)$ Yukawa coupling, $M$ is at the grand unification scale, which is appealing as $\nu_R$ naturally appears in the  SO(10) grand unification model.

 Meanwhile, it has recently been noticed that parameters in the EFT are severely constrained by quantum gravity, hence if we take quantum gravity effects into account appropriately, the naturalness criterion can change drastically \cite{Vafa:2005ui} (for reviews, see, e.g., \cite{Brennan:2017rbf, Palti:2019pca, vanBeest:2021lhn, Grana:2021zvf, Agmon:2022thq, VanRiet:2023pnx}). 
 A series of studies in this direction called `swampland program'  claims that some value of a parameter that seems to be natural when quantum gravity effects are completely ignored can be in fact unnatural, and even forbidden.
 Moreover, it has been shown that the Majorana neutrino mass conflicts with several `swampland conjectures', the conjectured   quantum gravity restrictions on the EFT \cite{Ibanez:2017kvh, Hamada:2017yji, Gonzalo:2018tpb, Gonzalo:2018dxi, Rudelius:2021oaz, Gonzalo:2021fma, Gonzalo:2021zsp, Casas:2024clw}.
 This can be argued by considering the compactification of the SM on a circle and observing the behavior of the radion potential in the 3-dimensional spacetime \cite{Arkani-Hamed:2007ryu}.
 At tree level, the radion potential  is proportional to the nonvanishing (4-dimensional) positive cosmological constant $\Lambda_4$ and exhibits a runaway behavior.
   This is corrected at 1-loop level by the Casimir energy coming from loops wrapping the circle, where the sign of the contribution from the fermions (bosons) is the same as (opposite to) the tree level potential.
  Suppose the lightest neutrino neutrino mass is 1) larger than $\Lambda_4^{1/4}$ such that the contribution of the neutrino is exponentially suppressed or 2) of the Majorana type such that the number of degrees of freedom is reduced by half   compared to that of the Dirac type.
  Then the 1-loop contribution of the neutrino which dominates the fermionic Casimir energy is not large enough to maintain the runaway behavior against that of massless bosons (the photon and the graviton) which tends to pull down the potential to the negative value.
  As a result, the potential is stabilized at the anti-de Sitter (AdS) vacuum.
  However, this may not be allowed by quantum gravity : according to the `AdS non-SUSY conjecture', the AdS vacuum cannot be stable unless it is supersymmetric \cite{Ooguri:2016pdq}.
 In addition, even though the transition among  AdS, Minkowski, and dS vacua seems to be continuous with respect to the parameters in the EFT, string theory suggests that it is not the case.
This is because the small value of the cosmological constant $\Lambda_{\rm cc}$ (in any dimension, and not necessarily positive) appearing in the continuous transition is realized in the asymptotic limit of the moduli space, where according to the distance conjecture \cite{Ooguri:2006in} the mass scale of a tower of states such as the Kaluza-Klein (KK) or the string tower becomes  extremely low, invalidating the EFT \cite{Lee:2019xtm, Lee:2019wij}.
Thus, we expect that in the vanishing limit of $\Lambda_{\rm cc}$, there exists a tower of states with the tower mass scale given by $|\Lambda_{\rm cc}|^\alpha$ for some positive value of $\alpha$, as claimed by the `AdS distance conjecture'  (see, e.g., \cite{Cribiori:2021gbf, Castellano:2021yye, Seo:2023ssl} for the similar version of the conjecture and also \cite{Seo:2023fuj} for the discussion on the discontinuity of the vacuum transition).
 It tells us that any (the Dirac as well as the Majorana) neutrino mass giving the tiny absolute value of the 3-dimensional cosmological constant cannot be allowed by string theory \cite{Gonzalo:2021fma, Gonzalo:2021zsp}.
 
 Similarly to the type I see-saw mechanism, the Dirac neutrino mass is realized by introducing the right-handed neutrino $\nu_R$ and adding the Yukawa term $(l\cdot h)\nu_R$ to the SM action :
 the difference arises from the absence of  the  Majorana mass term for  $\nu_R$.
 But even in this case, the Weinberg operator can be still written in the EFT action since it is consistent with gauge invariance of the SM.
 Indeed,  the Yukawa interaction $(l\cdot h)\nu_R$ is not an essential ingredient for the UV completion of the Weinberg operator.
 For example, in the type II see-saw mechanism the Weinberg operator is generated by integrating out the SU(2) triplet scalar which couples to $l$ and $h$ respectively \cite{Magg:1980ut, Cheng:1980qt, Lazarides:1980nt, Mohapatra:1980yp}.
 Therefore, it is not strange that the neutrino has both the Dirac and the Majorana mass simultaneously.
 Then since $\nu_R$ in the Dirac mass term prevents the number of the neutrino degrees of freedom from being reduced by half, if the Majorana neutrino mass from the Weinberg operator is much larger than the Dirac one, but there are sufficient degrees of freedom %(more precisely, at least $4$, i.e., the number of degrees of freedom for the single lightest Dirac neutrino) 
  whose masses are smaller than $\Lambda_4^{1/4}$, we may observe the Majorana nature of the neutrino without contradiction to the swampland constraints concerning the compactification of the SM on a circle (see also \cite{Ibanez:2017kvh}).
 This motivates us to investigate the (conjectured) quantum gravity constraints on the Majorana neutrino mass which do not rely on the compactification to the 3-dimensional spacetime.

 For this purpose, we use the fact that in the presence of a tower of states, their contributions   to the loop corrections of the relevant or the marginal operators through the irrelevant operators  are no longer negligible.
 To see this, suppose  a particle $\phi$ in the EFT couples to $N$ particle species  through the irrelevant operators of the same type, i.e., with the same coupling   which is suppressed by some positive power  of the mass scale $f$.
 Here $f$ is typically higher than the cutoff scale $\Lambda$.
 Then the total contribution of tower states to the loop correction of the relevant or the marginal operator containing $\phi$ through the irrelevant operators is proportional to $N(\Lambda/f)^p$ where $p$ is positive.
 Therefore, even if $(\Lambda/f)^p<1$, it is not suppressed provided $N \gtrsim (f/\Lambda)^p$, resulting in the breakdown of the perturbativity.
  Since states in a tower like the KK or the string tower couple to other particles through interactions of the same form, the sum of their contributions easily dominates the loop correction as described above, in particular, when the tower mass scale becomes very low  as predicted by the distance conjecture \cite{Ooguri:2006in}. 
If we require the EFT to be perturbative, one finds that   the cutoff scale must be lower than $\Lambda_{\rm sp, irr}$, the `species scale' associated with the irrelevant operator, satisfying  $N_{\rm sp, irr}(\Lambda_{\rm sp, irr}/f)^p \sim {\cal O}(1)$, where $N_{\rm sp, irr}$ is the number of particle species (with mass  below $\Lambda_{\rm sp, irr}$)  which couple to $\phi$  through the same form of the irrelevant operator.
Whereas the species scale has been typically defined for the gravitational interaction (in this case, $f=M_{\rm Pl}$, the Planck scale) \cite{Veneziano:2001ah, Dvali:2007hz, Dvali:2007wp, Dvali:2009ks, Dvali:2010vm, Dvali:2012uq}, we may define different species scales associated with various  irrelevant operators in the EFT.
For example, the axion couples to the gauge bosons or the fermions through the irrelevant operators with $f$ given by the Peccei-Quinn scale, thus the associated species scale can be found \cite{Reece:2024wrn, Seo:2024zzs}.
Such a species scale associated with the irrelevant operator is required to be lower than the gravitational species scale since it is typically obtained without considering the gravitational effects.
In the case of the axion, this  condition is equivalent to the bound on the Peccei-Quinn scale predicted by the weak gravity conjecture, $(8\pi^2/g^2)f \lesssim M_{\rm Pl}$ where  $g$ is some gauge coupling  \cite{Arkani-Hamed:2006emk}.
In our case of the neutrino, the species scale associated with the Weinberg operator which we will call the `Majorana species scale' can be defined as well.
Imposing the condition that this scale is lower than the gravitational species scale, we obtain  the upper bound on $M$,  which gives the conjectured lower bound on the Majorana neutrino mass.
 If we impose other swampland conjectures in addition, different lower bound on the Majorana neutrino mass can be found as well.

 To see the discussions sketched so far in detail, in  Sec. \ref{Sec:basics}, we first show  how a tower of states associated with the neutrino contributes to the loop corrections of the relevant or the marginal operators containing the Higgs through the  Weinberg operator.
  From this, we can define the Majorana species scale.
  In Sec. \ref{Sec:MajSp}, the possible lower bounds on the Majorana neutrino mass are obtained by considering several swampland conjectures, in particular the condition that the Majorana species scale is lower than the gravitational species scale.
  After discussing physical implications of our results, we conclude.
Throughout this article,  we use the two-component spinor notation, since it not only is fundamental in light of the representation of the (little group of the) Lorentz group, but also distinguishes the mass term from the kinetic term in the two-point correlator more manifestly.
For details, we refer the reader to   \cite{Dreiner:2008tw}, the version of which adopting the $(-+++)$ metric convention we follow can be found in \cite{tcsp}.

 \section{Majorana species scale}
 \label{Sec:basics}

In this article, we consider the interaction between a  spinor $\xi$ (``neutrino") and a real scalar $\phi$ (``Higgs") through the dimension-5 operator $\phi^2\xi\xi/(4M)$ (``Weinberg operator").
 In the presence of a tower of states associated with the neutrino, states in a tower couple to the Higgs in the same way, resulting in the sizeable contributions to the loop corrections of the Higgs quartic coupling and the  mass.
 The perturbativity of these corrections requires that the cutoff scale cannot exceed the Majorana species scale.

\subsection{A tower of states associated with the  neutrino}

We begin our discussion by observing the coupling between the Higgs and a tower of states associated with the neutrino through the Weinberg operator.
For this purpose, we take  the KK tower as an example, from which one is convinced  that   states in a tower couple to   the Higgs in the same way.
Consider the compactification of the 5-dimensional spacetime on a circle of radius $R$.
In this case, the 4-dimensional spinors with the KK mass of the Majorana type are obtained from the Dirac spinor $\Psi$, which has the same form in both the 4- and the 5-dimensional spacetime \cite{Arkani-Hamed:1998wuz} :
\dis{\Psi=\left(
\begin{array}{c}
\Xi_\alpha(x, y) \\
{\mathrm{H}^\dagger}^{\dot\alpha}(x, y)\\
\end{array}\right),
\quad\quad
\overline{\Psi}=\Big( \mathrm{H}^\alpha (x, y), \Xi^\dagger_{\dot\alpha}(x, y) \Big).}
Here $x$ and $y$ denote the coordinates of the 4-dimensional spacetime and the coordinate of the 5th dimension, respectively.
The 4-dimensional action can be obtained by the dimensional reduction of the 5-dimensional one, in which the spinors $\Xi$ and $\mathrm{H}$ (of the mass dimension $2$) and the real scalar $\Phi$ (of the mass dimension $3/2$) are expanded as  
\dis{&\Xi^\alpha(x, y)=\frac{1}{\sqrt{\pi R}}\Big[\frac{\xi^\alpha(x)}{\sqrt2}+\sum_{n \in \mathbb{Z}_{> 0}} \xi_n^\alpha (x) \cos \Big(\frac{n}{R}y\Big)\Big],  \quad
\mathrm {H}^\alpha=\frac{1}{\sqrt{\pi R}} \sum_{n \in \mathbb{Z}_{> 0}} \xi_n^\alpha (x) \sin \Big(\frac{n}{R}y\Big),
\\
&\Phi(x, y)=\frac{1}{\sqrt{2\pi R}}\sum_{n \in \mathbb{Z}} \phi_n (x) e^{i\frac{n}{R}y}.\label{eq:modeexp}}
Here the zero modes of the spinor and the scalar will be identified with the neutrino $\xi(x)$   and the Higgs $\phi(x)$, respectively.
Moreover, $\Xi$ ($\mathrm{H}$) is (anti-)symmetric under $y\to -y$, since otherwise the nonzero KK mass is not realized due to the cancellation between the positive and the negative KK mass terms (see, e.g., \cite{Pilaftsis:1999jk}).
\footnote{More concretely, if we expand $\Xi^\alpha$ in the same way as $\Phi$ and set  $H_\alpha=\Xi_\alpha$, there is a cancellation between $(n/R)\xi_{-n}\xi_n$ and $(-n/R)\xi_{n}\xi_{-n}$ since 
\dis{\xi_{-n}\xi_n =\xi_{-n}^\alpha {\xi_{n}}_\alpha =\epsilon^{\alpha\beta}{\xi_{-n}}_\beta {\xi_{n}}_\alpha=(-\epsilon^{\beta\alpha})(-{\xi_{n}}_\alpha{\xi_{-n}}_\beta )=\xi_n^\beta {\xi_{-n}}_\beta =\xi_n\xi_{-n},} 
where we use the facts that $\epsilon^{\alpha\beta}$ is antisymmetric and $\xi_n$ are fermionic.}
The reality of $\Phi$ gives the condition $\phi_n^\dagger = \phi_{-n}$, implying that the zero mode $\phi$ is a real scalar.
Noting that the Dirac matrices in 5 dimensions $\Gamma^M$ are given by those in 4 dimensions $\gamma^\mu$ and $i\gamma_5=- \gamma^0\gamma^1\gamma^2\gamma^3$, we obtain the free action for the KK modes of the Higgs and the neutrino
\footnote{Following \cite{tcsp}, explicit forms of the Dirac matrices in 4 dimensions and $\gamma_5$ are given by 
\dis{\gamma^\mu=\left(
\begin{array}{cc}
0 & \sigma^\mu_{\alpha{\dot\beta}} \\
{\overline{\sigma}^\mu}^{{\dot\alpha}\beta} & 0\\
\end{array}\right),
\quad\quad
\gamma_5 = \left(
\begin{array}{cc}
-\delta_\alpha^{~\beta} & 0\\
0 & \delta^{\dot\alpha}_{~{\dot\beta}}\\
\end{array}\right),}
respectively, and the 5-dimensional  Clifford algebra is given by $\{\Gamma^M, \Gamma^N\}=-2\eta^{MN}$.} : 
\dis{S=\int d^4 x &\int_0^{2\pi R} dy \Big[-\frac12 \partial_M \Phi \partial^M \Phi -\frac12 m^2\Phi^2 +\frac{i}{2}\overline{\Psi}\Gamma^M \partial_M\Psi\Big]
\\
=\int d^4x &\sum_{n \in \mathbb{Z}}\Big[-\frac12\partial_\mu \phi_{-n} \partial^\mu \phi_n-\frac12\Big(m^2+\frac{n^2}{R^2}\Big)\phi_{-n}\phi_n\Big]
\\
&+i \xi^\dagger  \overline{\sigma}^\mu \partial_\mu \xi  + \sum_{n \in \mathbb{Z}_{> 0}}\Big[i \xi^\dagger_n \overline{\sigma}^\mu \partial_\mu \xi_n-\frac12 \frac{n}{R}(\xi_n \xi_n+\xi_n^\dagger \xi_n^\dagger)\Big],\label{eq:free}}
where $\xi_n \xi_n=\xi_n^\alpha {\xi_n}_\alpha$ and $\xi_n^\dagger \xi_n^\dagger={\xi_n^\dagger}_{\dot\alpha} {\xi_n^\dagger}^{\dot \alpha}$.
 %%%%%%%%%%%%%%%%%%%%%%%%%%%%%%%%%%%%%%%%%%%%%%%%%%%%%%%%%%%%%%%%%%%%%%%%%%%%%%%%%%%%%%%%
 \begin{figure}[!t]
 \begin{center}
 \includegraphics[width=0.50\textwidth]{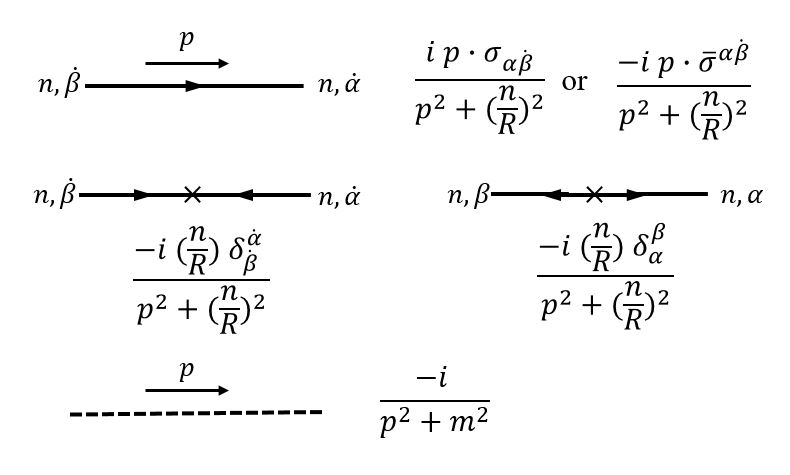}
\center{$(a)$}
 \\
 \includegraphics[width=0.70\textwidth]{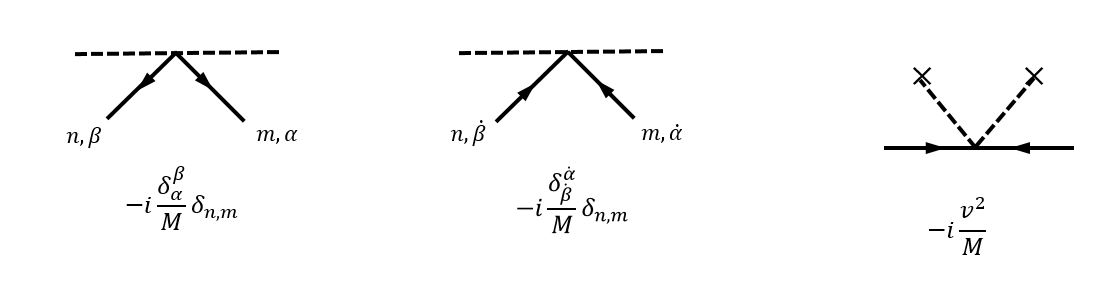}
\center{$(b)$} 
 \end{center}
\caption{$(a)$ : Propagators for $\xi_n$, $\xi_n^\dagger$ with $n\in \mathbb{Z}_{\geq 0}$ (arrow lines) and $\phi$ (dotted line). The arrow in the $\xi_n$($\xi_n^\dagger$) line toward(runaway from) the undotted(dotted) index represents  the left(right)-handed spinor.
The KK mass insertions in the propagators of $\xi_n$ and $\xi_n^\dagger$ are indicated by the cross.
$(b)$ : Interaction vertices for the Weinberg operator and the mass of the neutrino (i.e., zero mode $\xi$) generated by the Weinberg operator. The cross indicates that the Higgs VEV is inserted.}
\label{Fig:feynman}
\end{figure}
%%%%%%%%%%%%%%%%%%%%%%%%%%%%%%%%%%%%%%%%%%%%%%%%%%%%%%%%%%%%%%%%%%%%%%
On the other hand, the Weinberg operator is obtained from the dimensional reduction of the operator $\lambda \Phi^2 \overline{\Psi^c}\Psi/M_*^2$, where $\overline{\Psi^c}=(\Xi^\alpha, \mathrm{H}_{\dot\alpha}^\dagger)$ : 
\dis{ \int_0^{2\pi R}dy &\frac{\lambda}{M_*^2} \Phi^2 \overline{\Psi^c}\Psi
\\
=\frac{\lambda}{2\pi R M_*^2}& \Big[\sum_n \xi\xi \phi_n \phi_{-n}+\sqrt2 \sum_{n,p}\xi\xi_n \phi_p (\phi_{-n-p}+ \phi_{n-p})
\\
&+\frac12\sum_{m,n,p}\xi_m\xi_n \phi_p (\phi_{-m-n-p}+\phi_{m+n-p}+\phi_{-m+n+p}+\phi_{m-n-p})
\\
&+\frac12\sum_{m,n,p}\xi^\dagger_m\xi^\dagger_n \phi_p (-\phi_{-m-n-p}-\phi_{m+n-p}+\phi_{-m+n+p}+\phi_{m-n-p})\Big].}
Then we can define the mass scale $M=\pi R M_*^2/(2\lambda)$ such that the Weinberg operator for the zero modes $\xi$ and $\phi$ is written as $\xi\xi \phi^2/(4M)$.
The interactions associated with $\phi$ we are mainly interested in are given by
\dis{ \frac{1}{4M}& \Big[  \xi\xi \phi^2
+\sum_{n\in \mathbb{Z}_{>0}} (\xi_n\xi_n + \xi^\dagger_n\xi^\dagger_n) \phi^2\Big].}
The Feynman rules for the interactions above as well as the propagators are depicted in Fig. \ref{Fig:feynman}.
When the Higgs $\phi$ acquires the VEV $v$, the Weinberg operator generates the tree level Majorana mass of the neutrino $\xi$ (that is, the zero mode) given by $v^2/M$, which is also shown in  Fig. \ref{Fig:feynman} (b).
The Majorana mass of $\xi_n$ ($n>0$) is  the sum of that generated by the Weinberg operator and the KK mass  obtained from the dimensional reduction.
If the Higgs is localized on the three-brane, the KK modes of the Higgs do not exist.

The KK tower of the neutrino we have discussed so far corresponds to an example showing that states in a tower associated with the  neutrino couple  to   the Higgs  in the same way : all the interactions have the form of the Weinberg operator and the couplings are commonly given by $1/M$.
Meanwhile, it was claimed that in string theory towers which become light in the asymptotic limit of the moduli space are either the KK or  the string tower \cite{Lee:2019xtm, Lee:2019wij}.
From this we expect that the Higgs also couples to a string tower associated with the  neutrino.
For example, both the Higgs and the   neutrino can be  zero modes of the open string connecting intersecting D-branes or D-branes on top of each other, then  the Weinberg operators can be found from the quartic vertex for four open strings (two for the Higgs and remaining two for the string tower associated with the   neutrino).
%In the EFT, the dimensional consistency implies that the corresponding interaction may be identified with the Weinberg operator.

\subsection{Majorana species scale from loop corrections}
\label{Subsec:MjLambda} 

We now investigate the loop contribution of the tower of states  to the relevant/marginal operators containing the Higgs, namely, the Higgs quartic coupling and the mass, through the Weinberg operator.
But before going into detail, we first describe the generic situation.
Suppose some particle $\xi$ (neutrino in our case) interacts with another particle $\phi$ (the Higgs) through the irrelevant operator ${\cal O}_{\rm irr}[\xi, \phi]$ (the Weinberg operator), the mass scale of which is given by $f$ ($f=M$ for the Weinberg operator).
  Since  $f$ is typically higher than the cutoff scale $\Lambda$, one  expects that the contributions of $\xi$ to the loop corrections of the relevant/marginal operators containing $\phi$ through ${\cal O}_{\rm irr}[\xi, \phi]$ are suppressed by $(\Lambda/f)^p$ for some positive number $p$.
  That is, the irrelevant operator does not play the crucial role in the renormalization of the relevant/marginal operators.
However, the situation changes when a large number of states $\xi_i$ ($i=1,\cdots, N$ with $\xi_1=\xi$) interact with $\phi$ through   the irrelevant operators ${\cal O}_{\rm irr}[\xi_i, \phi]$ of the same form, which can be realized by the presence of a tower of states associated with $\xi$.
In this case, the loop contribution of $\xi_i$s through a set of irrelevant operators $\{{\cal O}_{\rm irr}[\xi_i, \phi]\}$  is proportional to $N (\Lambda/f)^p$ and when  it becomes ${\cal O}(1)$, not only does $\{{\cal O}_{\rm irr}[\xi_i, \phi]\}$  give the main contribution to the loop correction, but also the perturbativity breaks down. 
It means that the low energy EFT in  which the renormalization of the parameters is well controlled by the dominance of the contribution of the relevant/marginal operators over that of the irrelevant ones (in the sense of \cite{Polchinski:1983gv}) is no longer reliable.
\footnote{In other words, the macroscopic theory becomes sensitive to the microscopic theory which is encoded in the irrelevant operator.}
Therefore, the cutoff scale for the EFT must be lower than the `species scale' associated with ${\cal O}_{\rm irr}$ given by $\Lambda_{\rm sp, irr}=f/N_{\rm sp, irr}^{1/p}$, where $N_{\rm sp, irr}$ is the `species number', the number of $\xi_i$s below $\Lambda_{\rm sp, irr}$.
Whereas the species scale is originally defined in the context of the gravitational interaction  where $f=M_{\rm Pl}$ and $p=2$ \cite{Veneziano:2001ah, Dvali:2007hz, Dvali:2007wp, Dvali:2009ks, Dvali:2010vm, Dvali:2012uq}, it can be considered for any type of  interaction coming from the irrelevant operator such as the coupling between the axion and the gauge boson given by $\theta F\wedge F$ \cite{Reece:2024wrn, Seo:2024zzs} or the Weinberg operator for the   neutrino.

 %%%%%%%%%%%%%%%%%%%%%%%%%%%%%%%%%%%%%%%%%%%%%%%%%%%%%%%%%%%%%%%%%%%%%%%%%%%%%%%%%%%%%%%%
 \begin{figure}[!t]
 \begin{center}
 \includegraphics[width=0.60\textwidth]{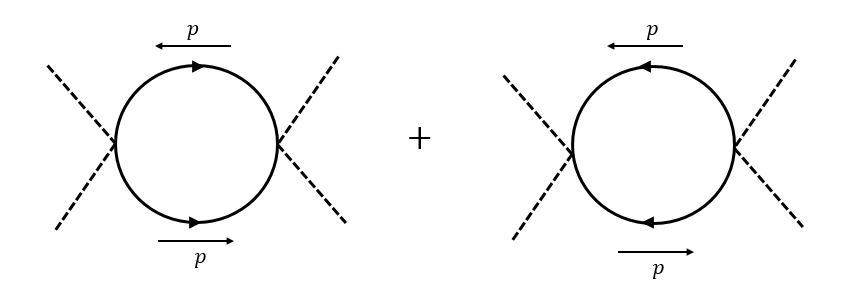}
 \end{center}
\caption{The $s$-channel diagram for the one-loop contribution of the tower states  to the Higgs quartic coupling through the Weinberg operator.}
\label{Fig:Higgsquartic}
\end{figure}
%%%%%%%%%%%%%%%%%%%%%%%%%%%%%%%%%%%%%%%%%%%%%%%%%%%%%%%%%%%%%%%%%%%%%%

Motivated by  observations above, let us  consider the loop contribution of the tower states to the Higgs quartic coupling through the Weinberg operator, the $s$-channel diagram of which is shown in Fig. \ref{Fig:Higgsquartic}.
There will be additional contributions from the crossed channels, i.e., the $t$- and the $u$-channels. 
Referring to an example of the KK tower  in the previous section, one finds that %when the heaviest tower state mass, i.e., the mass just below the cutoff scale $\Lambda$, is much smaller than $\Lambda$,   
 the leading term is  estimated as 
\dis{-i\delta \lambda\sim -\Big(\frac{-i}{M}\Big)^2 N \int \frac{d^4 p}{(2\pi)^4}\frac{{\rm Tr}\big[(-i p\cdot\overline{\sigma})(i(-p)\cdot\sigma)\big]}{(p^2)^2} \sim -\frac{i}{8\pi^2}N\Big(\frac{\Lambda }{M }\Big)^2, \label{eq:loop1} }
up to some ${\cal O}(1)$ constant counting the number of possible channels.
 \footnote{For more correct expression, one may calculate the loop integral with $p^2$ in the denominator replaced by $p^2+(n/R)^2$ and $N$ replaced by the summation over the KK modes labelled by $n$.
 Then $\Lambda^2$ in the RHS is replaced by
 \dis{\sum_n\frac{\Lambda^4+2 \Lambda^2 (n/R)^2-2(n/R)^2(\Lambda^2+(n/R)^2)\log[\frac{\Lambda^2+(n/R)^2}{(n/R)^2}]}{  \Lambda^2+(n/R)^2 }.}
 Here we consider the case where the KK mass scale $1/R$ is sufficiently large (but not too large to keep $N \gg 1$) such that the summation is dominated by the KK modes well below $\Lambda$ ($n \ll R\Lambda$), then the  estimation given by \eqref{eq:loop1} becomes correct.
 Whereas it is corrected by the KK modes with   $ n \sim R\Lambda$, so far as they are not too populated, their contributions are estimated to be at most ${\cal O}(1)\Lambda^2$, where the ${\cal }O(1)$ coefficient is typically suppressed compare to $N$.
 Then the expression in \eqref{eq:loop1} is a good estimation of the loop integral.  }
Thus for this contribution    to be suppressed, $\Lambda$ is required to be lower than the `Majorana species scale', the species scale associated with the Weinberg operator, given by
\dis{\Lambda_{{\rm sp},\nu} \simeq \frac{M}{\sqrt{N_{{\rm sp},\nu}}},\label{eq:Majspes}}
where the corresponding species number is denoted by $N_{{\rm sp},\nu}$.

 %%%%%%%%%%%%%%%%%%%%%%%%%%%%%%%%%%%%%%%%%%%%%%%%%%%%%%%%%%%%%%%%%%%%%%%%%%%%%%%%%%%%%%%%
 \begin{figure}[!t]
 \begin{center}
 \includegraphics[width=0.70\textwidth]{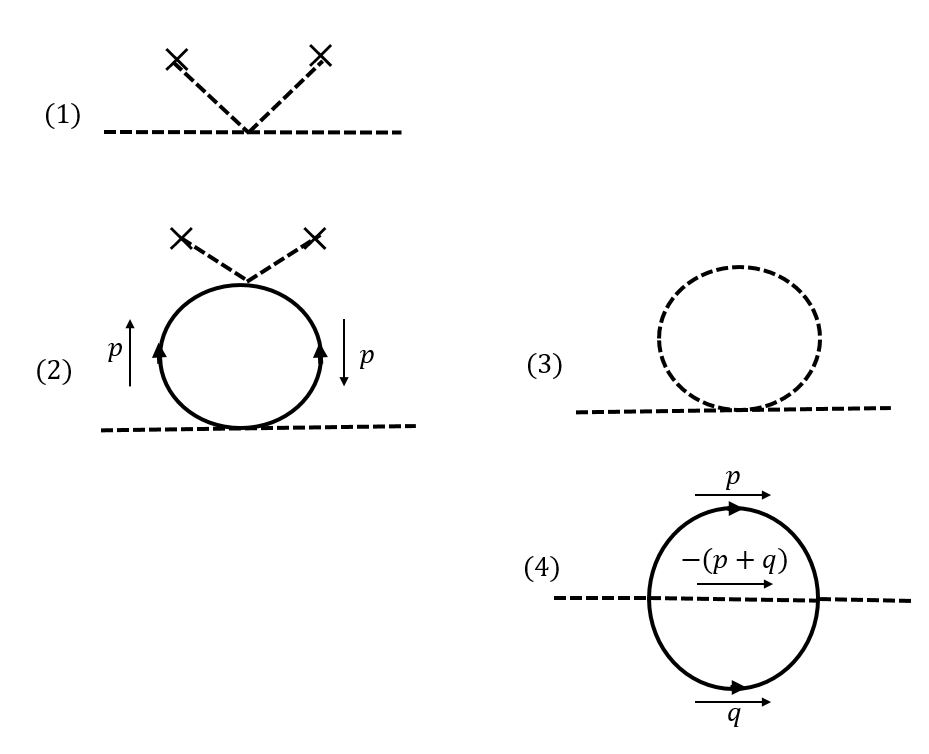}
 \end{center}
\caption{Feynman diagrams associated with the corrected Higgs mass up to  two-loop order, generated by the Higgs quartic coupling and the Weinberg operator. 
The cross indicates the insertion of the Higgs VEV.  }
\label{Fig:Higgsmass}
\end{figure}
%%%%%%%%%%%%%%%%%%%%%%%%%%%%%%%%%%%%%%%%%%%%%%%%%%%%%%%%%%%%%%%%%%%%%%

 The same result can be drawn by considering the loop contributions of the tower states  to the Higgs mass through the Weinberg operator.
 If we restrict the allowed interactions to the Higgs quartic coupling and the Weinberg operator, relevant Feynman diagrams for the corrected Higgs mass up to two-loop order are given by Fig. \ref{Fig:Higgsmass} (for the fermion loops, the same diagrams with the helicity reversed must be included).
 While the Higgs quartic coupling allows a diagram in the form of Fig. \ref{Fig:Higgsmass} (4) with all the internal lines are given by the Higgs propagators, we omitted this since it is suppressed compared to Fig. \ref{Fig:Higgsmass} (1) and   (3).
 The extension of the discussion including the Yukawa coupling between the Higgs and other SM fermions is straightforward.
 Now the tree-level diagram (Fig. \ref{Fig:Higgsmass} (1)) and the one-loop diagram in Fig.  \ref{Fig:Higgsmass} (2) are combined to give the corrected Higgs mass raised by the corrected Higgs quartic coupling, $m^2=(\lambda+\delta\lambda)v^2$, and from \eqref{eq:loop1} we obtain
  \dis{\delta m^2_{(2)} \equiv  \delta\lambda \times v^2\sim \frac{\lambda}{8\pi^2}N \Big(\frac{\Lambda }{M }\Big)^2 v^2.}
 For the Weinberg operator  to decouple from the correction to the Higgs mass, we again require that $\lambda v^2 \gg \delta m^2_{(2)}$, resulting in the upper bound on the cutoff scale given by \eqref{eq:Majspes}.
 On the other hand, the Higgs mass also can be corrected by diagrams  Fig. \ref{Fig:Higgsmass} (3) and (4) which are not relevant to the correction to $\lambda$.
 The one-loop diagram  Fig. \ref{Fig:Higgsmass}  (3) is generated by the Higgs quartic coupling and given by $\lambda\Lambda^2$ times the loop factor.
 Whereas the KK modes of the Higgs  also contribute to the loop, they may be absent if the Higgs is localized on the three-brane.
 In this case, the two-loop diagram generated by the Weinberg operator given by  Fig. \ref{Fig:Higgsmass}  (4), or equivalently, 
 \dis{\delta m^2_{(4)} \sim - \Big(\frac{-i}{M}\Big)^2 N \int\frac{d^4 p d^4 q}{(2\pi)^8} \frac{{\rm Tr}\big[(-i p\cdot\overline{\sigma})(iq\cdot\sigma)\big]}{p^2 q^2 (p+q)^2} \sim \frac{1}{(8\pi^2)^2}N \Big(\frac{\Lambda }{M }\Big)^2 \Lambda^2,} 
  is suppressed compared to the one-loop diagram  Fig. \ref{Fig:Higgsmass}  (3) provided $\Lambda$ is below the cutoff scale given by \eqref{eq:Majspes}.
 Therefore, \eqref{eq:Majspes} is a reasonable definition of the Majorana species scale.
 \footnote{Whereas the two-loop diagram $\delta m^2_{(4)}$ is much larger than the one-loop diagram $\delta m^2_{(2)}$ as $\Lambda \gg v$,   they can be treated separately since $\delta m^2_{(2)}=\delta \lambda v^2$  is absorbed to the renormalized  Higgs quartic coupling $\lambda$. }
 
 Explicit values of $\Lambda_{{\rm sp},\nu}$ and $N_{{\rm sp},\nu}$ depend on the model.
 To see this, let us express the spectrum of states in a tower as $m_n=n^{1/p} m_t$, where $m_t$ is a tower mass scale like the KK or the string mass scale.
 The positive integers $n$ and $p$ represent the step of the state in the tower and the number of states with identical mass gap, respectively \cite{Castellano:2021mmx}.
 For example, for the KK tower, $m_t$ is given by the KK mass scale and $p$ is the number of towers of the identical value of $m_t$. 
 On the other hand, for the string tower where $m_t$ is given by the string mass scale (or equivalently, the inverse of the string length $\ell_s^{-1}$), $m_n\sim \sqrt{n} m_t$ but the degeneracy grows exponentially as $e^{\sqrt{n}}$ \cite{Kani:1989im}, which gives $p=\infty$.
 In any case, $N_{{\rm sp},\nu}=(\Lambda_{{\rm sp},\nu}/m_t)^p$ is satisfied, and combining this with \eqref{eq:Majspes} we obtain
 \dis{\Lambda_{{\rm sp},\nu}\sim M^{\frac{2}{p+2}}m_t^{\frac{p}{p+2}},\quad\quad N_{{\rm sp},\nu}\sim \Big(\frac{M}{m_t}\Big)^{\frac{2p}{p+2}}.\label{eq:Sp1}}
 We note that for the string tower, the relations become  simple : $\Lambda_{{\rm sp},\nu} \sim \ell_s^{-1}$ and $N_{{\rm sp},\nu}\sim (M\ell_s)^2$.
 In the presence of multiple towers with different tower mass scales $m_{t_i}$ ($i=1, 2, \cdots$), the mass of the tower state is written as 
 \dis{ m^2_{\{ n_i\}}=\sum_i n_i^{2/p_i} m_{t_i}^2.\label{eq:towermass}}
 In this case, the Majorana species number and scale are given by
\dis{N_{{\rm sp},\nu}\sim \prod_i \Big(\frac{\Lambda_{{\rm sp},\nu}}{m_{t_i}}\Big)^{p_i}=\frac{M^{\frac{2p}{p+2}}}{\prod_i m_{t_i}^{\frac{2p_i}{p+2}}},\quad\quad  \Lambda_{{\rm sp},\nu}\sim M^{\frac{2}{p+2}} \prod_i m_{t_i}^{\frac{p_i}{p+2}}, \label{eq:Sp2} }
respectively \cite{Castellano:2021mmx, Castellano:2022bvr, Seo:2023xsb}, where $p=\sum_i p_i$.

 \section{Bound on the Majorana neutrino mass}
 \label{Sec:MajSp} 
 
As discussed at the beginning of Sec. \ref{Subsec:MjLambda}, the species scale can be defined for any interaction  coming from the irrelevant operator.
In particular, since every gravitational interaction in the EFT comes from the irrelevant operator, we can find the gravitational species scale given by
\dis{\Lambda_{\rm sp} =\frac{M_{\rm Pl}}{\sqrt{N_{\rm sp}}}.}
This can be easily obtained by imposing the perturbativity on the loop correction of the graviton propagator where all the states with mass below $\Lambda_{\rm sp}$ contribute  (the total number of which is given by the `(gravitational) species number' $N_{\rm sp}$).
Explicit expressions of $\Lambda_{\rm sp}$ and $N_{\rm sp}$ are the same as \eqref{eq:Sp1} or \eqref{eq:Sp2}, with $\Lambda_{{\rm sp}, \nu}$, $N_{{\rm sp}, \nu}$, and $M$ replaced by $\Lambda_{\rm sp}$, $N_{\rm sp}$, and $M_{\rm Pl}$, respectively.
Then $\Lambda_{\rm sp}$ is interpreted as the cutoff scale above which quantum gravity effects become manifest.

We now impose that any species scale $\Lambda_{\rm sp, irr}$ associated with the irrelevant operator is lower than the gravitational species scale.
This requirement is  motivated by the fact that quantum gravity is the most fundamental microscopic theory, and moreover, in obtaining $\Lambda_{\rm sp, irr}$, as can be noticed from the case of  the Majorana species scale $\Lambda_{{\rm sp},\nu}$, any  quantum gravity effect is not taken into account.
Our assumption here is that for the scale below which quantum gravity effects are negligible, we take $\Lambda_{\rm sp}$ rather than $M_{\rm Pl}$, requiring that gravity is weakly coupled in the EFT (even though it is not supported by the rigorous consistency condition, it is basically assumed in discussions concerning the species scale).
Indeed, it turns out that if we impose the same requirement on the species scale for the axion  which is obtained by considering the interaction between the axion and the gauge bosons through the irrelevant operator $\theta F\wedge F$, the axionic weak gravity conjecture bound $(8\pi^2/g^2)f \lesssim M_{\rm Pl}$ appears \cite{Reece:2024wrn, Seo:2024zzs}.
For the Majorana species scale, the condition $\Lambda_{{\rm sp},\nu} \lesssim \Lambda_{\rm sp}$ leads to the upper bound on the scale of the Weinberg operator $M$ given by
\dis{M \lesssim \sqrt{\frac{N_{{\rm sp},\nu}}{N_{\rm sp}}}M_{\rm Pl}.\label{eq:MvsMpl}}
We note that whereas $N_{\rm sp}$ counts the total number of  states   with mass below $\Lambda_{\rm sp}$, $N_{{\rm sp},\nu}$ counts the number of states in towers associated with the  neutrino exclusively.
Moreover, the mass of the state counted in $N_{{\rm sp},\nu}$ is lower than  $\Lambda_{{\rm sp},\nu}$ which is below $\Lambda_{\rm sp}$.
These indicate that the ratio $N_{{\rm sp},\nu}/N_{\rm sp}$ is smaller than one, therefore $M<M_{\rm Pl}$, i.e., the UV completion of the Weinberg operator appears at the mass scale below $M_{\rm Pl}$, as typically assumed in the particle physics models for the Majorana neutrino mass.
The nontrivial feature of this bound is that the hierarchy between $M$ and $M_{\rm Pl}$ is determined by the number of species.
Indeed,  if  $N_{{\rm sp},\nu}/N_{\rm sp}$ is too small such that $M$ is far below $M_{\rm Pl}$, the Majorana neutrino mass can be heavier than the observational bound.
This sets a constraint on the number of  low energy degrees of freedom.
For instance, since the gravitational interaction determining $N_{\rm sp}$ is not restricted to the SM sector, the size of the hidden sector degrees of freedom cannot be arbitrarily large.
Moreover, if the towers associated with the neutrino are just given by the KK towers, $N_{{\rm sp},\nu}$ is typically much smaller than $N_{\rm sp}$, thus $M$ is much lower than $M_{\rm Pl}$, and as we will see, this assumption gives rise to some interesting   phenomenological predictions.
 On the other hand, when the string tower is taken into account, since the number of  states in the string tower is exponentially large,  the string tower gives the dominant contribution to towers of states associated with the neutrino in size.
 Let one of towers, say, a tower labelled by $i=1$, be given by the string tower.
Then  from \eqref{eq:Sp2}, with $m_{t_1}=\ell_s^{-1}$ and $p \simeq p_1 =\infty$, we obtain  $\Lambda_{{\rm sp},\nu} \sim \ell_s^{-1}$ and $N_{{\rm sp},\nu}\sim (M\ell_s)^2$.
As an exponentially large number of string excitations appear  below $\Lambda_{{\rm sp},\nu}$ hence $\Lambda_{\rm sp}$,  the relations $\Lambda_{\rm sp}\sim \ell_s^{-1} \sim \Lambda_{{\rm sp},\nu}$  and $N_{\rm sp} \sim (M_{\rm Pl}\ell_s)^2$ are satisfied as well, resulting in the saturation of the inequality \eqref{eq:MvsMpl} ($M\sim \sqrt{N_{{\rm sp},\nu}/N_{\rm sp}}M_{\rm pl}$).
We also note that  $N_{{\rm sp},\nu}$ in this case mainly counts all the fermionic excitations of the string whose zero modes include the neutrino.
If only a few types of the string (distinguished by, for example,  a pair of branes connected by the string) are sufficient to describe the nature,  $N_{{\rm sp},\nu} $ may not be much suppressed compared to $N_{\rm sp}$ thus $M$ can be close to $M_{\rm Pl}$.

 Before discussing phenomenological consequences of our assumed condition $N_{{\rm sp},\nu} \lesssim N_{\rm sp}$, we comment on another fundamental scale associated with the neutrino.
 \footnote{We are grateful to the referee for pointing out this.}
 In the presence of $p$ extra dimensions with volume $V_p$, the fundamental scale for the Weinberg operator is given by $M_*=(M/ V_p)^{\frac{1}{p+1}}$.
 For example, in the five-dimensional model in Sec. \ref{Sec:basics}, $p=1$ and $V_p=R$ give the correct relation $M\sim R M_*^2$.
 If the internal manifold is almost isotropic, we may estimate $M_*$ as $M^{\frac{1}{p+1}} m_{t}^{\frac{p}{p+1}}$ where $m_t$ is the KK mass scale.
 Comparing this with \eqref{eq:Sp1}, one finds that $M_*$ is different from $\Lambda_{{\rm sp},\nu}$. 
 For the five-dimensional model, these two scales are related by $\Lambda_{{\rm sp},\nu} \sim R^{1/3} M_*^{4/3} =(M_*/m_t)^{1/3}M_*$.
 Since we typically impose $M_*> m_t$ (thus the Weinberg operator is irrelevant even if we can see the extra dimension), $\Lambda_{{\rm sp},\nu}$ is well above $M_*$.
 On the other hand, for the gravitational interaction, the species scale for the KK modes is nothing more than the higher dimensional Planck scale.
 Such a  discrepancy originates from the dimension of the irrelevant operator : whereas the leading gravitational interaction is suppressed by $M_{\rm Pl}^2$, the dimension-$5$ Weinberg operator is suppressed by $M$.
 Indeed, if the dimension of Weinberg operator were $6$ such that it is suppressed by $M^2$, $\Lambda_{{\rm sp},\nu}$ and $M_*$ would coincide.
 Therefore, in the presence of extra dimensions, we can implement two scales : the scale obtained from the condition of perturbativity and the fundamental scale in view of the extra dimensions.
 
 In any case, in the EFT point of view, $\Lambda_{{\rm sp},\nu}$ is required to be similar to $M_*$ in size, since $M_*$ is a natural cutoff scale for the neutrino physics in the full $4+p$ dimensions.
This sets the bound on the size of the extra dimension, which corresponds to our phenomenological assumption for the validity of discussion.
To see this, suppose we fix the size of $M$, which will be determined by the observed neutrino mass, in terms of which we obtain
\dis{\frac{\Lambda_{{\rm sp},\nu}}{M_*}=\Big(\frac{M_*}{m_t}\Big)^{\frac{p}{p+2}}=\Big(\frac{M}{m_t}\Big)^{\frac{p}{(p+1)(p+2)}}.\label{eq:ratios}}
For a comparison, we recall that
\dis{M_*\sim \Big(\frac{M}{m_t}\Big)^{\frac{1}{p+1}}m_t,\quad\quad
\Lambda_{{\rm sp},\nu}\sim \Big(\frac{M}{m_t}\Big)^{\frac{2}{p+2}} m_t.}
When $p=1$, the ratio $\Lambda_{{\rm sp},\nu}/M_*$ does not exceed ${\cal O}(10)$ provided $M/m_t \lesssim 10^6$, which corresponds to $M_*/m_t \lesssim 10^3$ and $\Lambda_{{\rm sp},\nu}/m_t \lesssim 10^4$.
More hierarchy in $M/m_t$ makes $\Lambda_{{\rm sp},\nu}$ too far above $M_*$, so physics concerning $\Lambda_{{\rm sp},\nu}$ is not reliable.
Whereas it is more or less acceptable phenomenologically, there is more room for the larger number of extra dimensions.
This can be noticed from the last expression in \eqref{eq:ratios}, showing that the ratio $\Lambda_{{\rm sp},\nu}/M_*$ approaches to $1$ as $p$ increases.
This comes from the fact that, for a fixed value of $M$, $M_*$ gets closer to $m_t$ as $p$ increases.
Indeed, the extreme case of $p\to \infty$ is nothing more than the case of the string tower, where $\Lambda_{{\rm sp},\nu}$ as well as $M_*$ coincides with $m_t$, the string mass scale, due to the exponentially large number of states in a tower.
In the case of the KK tower, it is typical that $p$ is not infinitely large, which keeps the ratio $M_*/m_t$ sizeable.
For example, when $p=6$, the ratio $\Lambda_{{\rm sp},\nu}/M_*$ does not exceed ${\cal O}(10)$ provided $M/m_t \lesssim 10^9$, which corresponds to $M_*/m_t \lesssim 20$ and $\Lambda_{{\rm sp},\nu}/m_t \lesssim 200$.
While $M_*$ is quite close to $m_t$ in this case, if only two or three KK mass scales are relevant to the species scale, $M_*/m_t \sim 10^2$ is allowed, which is also acceptable phenomenologically.

The upper bound on $M$ given by \eqref{eq:MvsMpl} leads to the following lower bound on the Majorana neutrino mass : 
\dis{m_\nu =\frac{v^2}{M} \gtrsim  \sqrt{\frac{N_{\rm sp} }{N_{{\rm sp},\nu}}} \frac{v^2}{M_{\rm Pl}} \gtrsim \frac{v^2}{M_{\rm Pl}}.\label{eq:numassbound}}
Comparing with $M_{\rm Pl}=2.4\times 10^{18}$ GeV, the Higgs VEV $v=246$ GeV is suppressed as $v\sim 10^{-16}M_{\rm Pl}$ thus the rightmost term is given by $v^2/M_{\rm Pl} \sim 2.5\times 10^{-5} {\rm eV}$ $\sim10^{-32}M_{\rm Pl}$.
As well known from the estimation supporting the see-saw mechanism at the grand unification scale, it is about $10^2 - 10^3$ times smaller than the observational upper bound on the neutrino mass  given by $\sim (10^{-2}-10^{-3})$eV.
\footnote{ From the neutrino oscillation data the neutrino mass difference is measured as $\Delta m_\nu^2=10^{-3}-10^{-5}$ GeV$^2$ \cite{ParticleDataGroup:2024cfk} and the astrophysical bound shows that the sum of three neutrino masses in the SM is smaller than $0.1$ eV \cite{GAMBITCosmologyWorkgroup:2020rmf, DiValentino:2021hoh}.}
Indeed, the quantity in the middle, $\sqrt{N_{\rm sp}/N_{{\rm sp},\nu}}(v^2/M_{\rm Pl})$, coincides with this neutrino mass bound  if a factor $10^2 - 10^3$   leading to $M\sim (10^{-3}-10^{-2})M_{\rm Pl}$  is provided by the large ratio $\sqrt{N_{{\rm sp},\nu}/N_{\rm sp}}$.
To see this, we assume the simplest case in which all the towers have the same mass scale $m_t$ hence \eqref{eq:Sp1} can be used. 
Whereas the total number of degrees of freedom in the SM sector is  $\sim {\cal O}(10^2)$,  since there may be a dark sector,  the number of degrees of freedom in the absence of towers can be estimated as ${\cal O} (10^2-10^3)$.
For the KK tower, each of these degrees of freedom has its own tower, which leads to  $N_{\rm sp}\sim (10^2-10^3)(M_{\rm Pl}/m_t)^{\frac{2p}{p+2}}$.
\footnote{
Of course, this is just a rough estimation, since not all dark sector states may not have the KK tower :  the states localized on the brane do not have KK excitations, decreasing $N_{\rm sp}$.
}
On the other hand, in the SM, there are three generations of   the neutrino, or equivalently, 6 neutrino degrees of freedom  which couple to the Higgs through the Weinberg operator, leading to $N_{{\rm sp},\nu} \sim 6(M/m_t)^{\frac{2p}{p+2}}$.
Therefore, we obtain 
\dis{\sqrt{\frac{N_{\rm sp} }{N_{{\rm sp},\nu}}}  \sim (10-10^2) \Big(\frac{M_{\rm Pl}}{M}\Big)^{\frac{p}{p+2}}.}
From $1\leq p < \infty$, one finds that the above quantity becomes $10^2 - 10^3$ provided $M\sim (10^{-1}- 10^{-3})M_{\rm Pl}$, which is more or less consistent with $M \sim (10^{-3}-10^{-2})M_{\rm Pl}$ we started with.
More concretely, putting this back to the expression for the neutrino mass $m_\nu = v^2/M$, we obtain $m_\nu \sim (10^{-31}-10^{-29})M_{\rm Pl}$ or $(10^{-4}-10^{-1})$ eV.
Meanwhile, if the string mass scale is low enough that the string tower gives the dominant contribution to the towers  associated with the   neutrino below $\Lambda_{{\rm sp},\nu}$, $\sqrt{N_{{\rm sp},\nu}/N_{\rm sp}} \lesssim {\cal O}(1)$ can be satisfied such that $M$ becomes close to  $M_{\rm Pl}$, then the lower bound on the Majorana neutrino mass is not enhanced but remains suppressed compared to the observational bound by a factor $10^{-3}-10^{-2}$.

 On the other hand, applying the AdS non-SUSY conjecture to    compactification of the SM on a circle, one obtains the upper bound on the Dirac neutrino mass $m_{\nu, D} \lesssim \Lambda_4^{1/4}\sim \sqrt{H  M_{\rm Pl}}$,  where $H$ is the Hubble parameter satisfying $\Lambda_4=3 H^2 M_{\rm Pl}^2$ \cite{Ibanez:2017kvh, Hamada:2017yji, Rudelius:2021oaz, Gonzalo:2021fma, Gonzalo:2021zsp, Casas:2024clw}.
 While the Majorana neutrino mass is claimed to be ruled out by the same consideration, this assumes that  the neutrino mass is given by the Majorana type only, in which the right-handed neutrino $\nu_R$ does not appear in the EFT hence the number of degrees of freedom  is reduced by half compared to the case of the Dirac type.
 However, the SM gauge invariance allows us to write down the Dirac neutrino mass term and the Weinberg operator simultaneously in the EFT action. 
 In this case, the number of degrees of freedom is not reduced due to the existence of $\nu_R$,  so if the Majorana mass generated by the Weinberg operator is much larger than the  Dirac mass but there are at least four degrees of freedom whose mass is smaller than $\Lambda_4^{1/4}$, we may  observe the Majorana nature of neutrino without contradiction to the swampland constraints.
 \footnote{The possibility that the Majorana neutrino is allowed due to the presence of extra light fermionic degrees of freedom is already discussed in \cite{Ibanez:2017kvh}.}
 Then the bound \eqref{eq:numassbound}  corresponds to the lower bound on the neutrino mass, which  is complementary to the upper bound on the neutrino mass obtained by considering  compactification of the SM on a circle.
 As emphasized in the literatures, the upper bound on the neutrino mass $\Lambda_4^{1/4}$ well coincides with the observational bound.
 To see this, we recall that the observed value of the cosmological constant is given by $\Lambda_4 \sim (2\times10^{-3} {\rm eV})^4 \sim 10^{-120} M_{\rm Pl}^4$ or equivalently, $H\sim 10^{-60}M_{\rm Pl}$.
 Then $\Lambda_4^{1/4}$ can be rewritten as $\sqrt{M_{\rm Pl}^3 H}/M_{\rm Pl}$, and since $\sqrt{M_{\rm Pl}^3 H} \sim 10^2 v^2$, it  almost coincides with the observational bound  without  the ambiguity of a factor $10^2-10^3$ appearing in the lower bound on the Majorana neutrino mass given by \eqref{eq:numassbound}.
This in fact was the motivation to look for quantum gravity constraints relating the neutrino mass and $\Lambda_4$.

 Then one may wonder if we can find  quantum gravity argument relating the Majorana neutrino mass to $\Lambda_4$ explicitly.
In fact, string theory which motivates  most of swampland conjectures does not tell much about the relation between $\Lambda_4$ and parameters in the EFT.
It is because in the string theory framework,   de Sitter space does not seem to be a stable vacuum solution \cite{Dine:1985he, Danielsson:2018ztv, Obied:2018sgi} (see also \cite{Andriot:2018wzk, Garg:2018reu, Ooguri:2018wrx, Hebecker:2018vxz, Andriot:2018mav}).
Nevertheless, there is a conjectured lower bound on $v$ determined by $\Lambda_4$ called the  `Festina-Lente' bound \cite{Montero:2019ekk} (for related discussions, see, e.g., \cite{Montero:2021otb, Lee:2021cor,  Guidetti:2022xct, Montero:2022jrc, Hassan:2024pht}).
 It states that  when $\Lambda_4$ is positive, the mass of the U(1) charged particle $m$ must be bounded from below as $m^4/\Lambda_4 \gtrsim g^2/(8\pi^2)$ where $g$ is the U(1) gauge coupling.
The supporting argument of this conjectured bound relies not on string theory but on the cosmic censorship conjecture \cite{Penrose:1969pc}, which forbids the exposure of the singularity.
%\footnote{We note that the weak gravity conjecture also comes from the requirement that the charged black hole must be discharged without the exposure of the singularity.}
More concretely, suppose in dS space, the charged black hole close to the Nariai limit (where the black hole horizon and the cosmological horizon  coincide) is produced.
Then the Festina-Lente bound is obtained by requiring that when this black hole is discharged by emitting the charged particles, it should not become super-extremal.
Since the mass of the U(1) charged particles in the SM is determined by  $v$, the Festina-Lente bound eventually sets the lower bound on $v$ given by
\dis{v^2 \gtrsim \frac{g}{4\pi y_e^2}\Lambda_4^{1/2} \simeq  \frac{g}{4\pi y_e^2}H M_{\rm Pl}, \label{eq:FL}}
where $y_e$ is the smallest value of the Yukawa coupling for the U(1) charged particle, namely, the electron Yukawa coupling.
Combining this with \eqref{eq:numassbound}, we obtain
\dis{m_\nu =\frac{v^2}{M} \gtrsim  \sqrt{\frac{N_{\rm sp} }{N_{{\rm sp},\nu}}}  \frac{g}{4\pi y_e^2} H.}
This bound is not so stringent since $H\sim 10^{-60}M_{\rm Pl}$ is even more suppressed compared to $\Lambda_4^{1/4}\sim  10^{-30}M_{\rm Pl}$ : even though $y_e \sim 10^{-5}$ is  a tiny value, the rightmost term is just given by $\sqrt{N_{\rm sp}/N_{{\rm sp},\nu}}\times 10^{-50}M_{\rm Pl}$.
This is not so strange since the observed value of $v$   is much larger than the bound given by the r.h.s. of \eqref{eq:FL}.
%The situation does not change drastically even if we   consider the inflationary era when the value of $\Lambda_4$ (hence $H$) may be much larger than the current observed value. 
%The value of $\Lambda_4$ during the inflation is constrained by two facts.
%First, after the end of the inflation, the vacuum energy would be converted into radiation, so we expect that $\Lambda_4 \simeq (\pi^2/30)g_{\rm reh}T_{\rm reh}^4$ where $g_{\rm reh}$ is the effective degrees of freedom during reheating and $T_{\rm reh}^4$ is the reheating temperature.
%Combining this with the Festina-Lente bound, we obtain $v\gtrsim (g_{\rm reh}^{1/2} g/(4\pi y_0^2))^{1/2}T_{\rm reh}$ \cite{Lee:2021cor}.
%Second, since the thermal history of the universe after the Big Bang nucleosynthesis (BBN) well coincides with   observations,  we can require that $T_{\rm reh}$ is larger than the BBN scale $10$ MeV  corresponding to  $g_{\rm reh}=10.75$ (see, e.g., Ch. 3.3 of \cite{Kolb:1990vq}).
%Then the lower bound on $v$ during the inflation is as large as $10^4$ GeV, thus the Majorana neutrino mass during the inflation is larger than $v^2/M_{\rm Pl}\sim 0.1$ eV.
%This is still much suppressed compared to $\Lambda_4^{1/4} \sim 10$ MeV.

 We close this section with a remark that when the neutrino has both the Majorana mass ($m_M$) and the Dirac   mass ($m_D$),  the physically observable mass would be the eigenvalues of the mass matrix  which is written in the basis of the left- and the right-handed neutrino states as
 \dis{\left(
\begin{array}{cc}
m_M & m_D \\
m_D & 0\\
\end{array}\right).}
Eigenvalues are explicitly given by $\frac12 \big(m_M\pm \sqrt{m_M^2+4m_D^2}\big)$, then when $m_M\gg m_D$, they are approximately as $m_M+(m_D^2/m_M)$  and $m_D^2/m_M$, respectively, after eliminating the minus sign by the phase redefinition.
Comparing with the case of $m_D \gg m_M$, where the mass eigenvalues are approximated as $m_D\pm m_M$, one finds that the   mass eigenvalues are more hierarchical.
Then when the SM is compactified on a circle, the lightest ones give the dominant contribution to the 1-loop Casimir energy, and prevent the vacuum energy from being stabilized at the AdS vacuum. % (if their number of degrees of freedom  is larger than $4$).
We also note that the small Yukawa coupling in the Dirac mass term may be appealing since in this case $\nu_R$ hardly couples to the SM particles so $N_{\rm eff}$, the effective number of the neutrino species in cosmology, can be close to $3$, which is consistent with the observations \cite{Dolgov:2002wy}.

 \section{Conclusions}
\label{sec:conclusion}

In this article, we try to obtain swampland constraints on the Majorana neutrino mass generated by  the dimension-5 Weinberg operator.
In the presence of a tower of states associated with the  neutrino, states in the tower couple to the Higgs through the interactions of the same form, namely,  the Weinberg operator.
Then the contributions of the tower to the loop corrections of the Higgs quartic coupling and the mass become  sizeable, threatening the  perturbativity.
 To avoid this, the cutoff scale   must be lower than the `Majorana species scale', the species scale associated with the  Weinberg operator, and by requiring it to be lower than the gravitational species scale, we obtain the upper bound on the mass scale of the Weinberg operator.
 Then we find the lower bound on the Majorana neutrino mass given by $v^2/M_{\rm Pl}$ times a factor coming from the ratio between the gravitational species number  and the  neutrino species number.
 If this factor is of ${\cal O}(1)$, the bound can be suppressed by $10^{-3}-10^{-2}$ compared to the observational bound   $\Lambda_4^{1/4}$.
  We can also obtain the lower bound on the Majorana neutrino mass using the Festina-Lente bound, but it gives much less stringent bound, $m_\nu \gtrsim (g/y_e^2)H$. 
 
  Even if the neutrino mass is of the Dirac type at the renormalizable level, i.e., in terms of the relevant and the marginal operators, our discussion is still valid so far as the Weinberg operator is not forbidden.
  In this case,  the Majorana mass is   generated as an effect of the irrelevant operator, namely, the Weinberg operator.
 If the Majorana mass is much larger than the Dirac mass but still there are sufficient degrees of freedom with mass smaller than $\Lambda_4^{1/4}$, the observation of the Majorana nature of the neutrino may not contradict to the swampland constraints concerning the compactification of the SM on a circle, since the non-supersymmetric AdS vacuum can be prevented by the right handed neutrino appearing in the EFT.
Moreover, our consideration of the Weinberg operator provides the lower bound on the neutrino mass, which is complementary to the upper bound obtained by observing the compactification of the SM on a circle.

%\subsection*{Acknowledgements}

%

%\newpage

\appendix

%\section{}
%\label{App:}

\renewcommand{\theequation}{\Alph{section}.\arabic{equation}}

%\section{Uncertainty for the infrared modes}
%\label{app:IRuncert}
%\setcounter{equation}{0}

\end{document}